\documentclass[reprint,aip,pop,graphicx,numerical,amsfonts,amsmath,amssymb,twocolumn]{revtex4-1}

\usepackage{bm}
\usepackage[dvips]{graphicx}
\usepackage{lineno}
\usepackage{color}
\usepackage{soul}
\usepackage{booktabs}
\usepackage{multirow}
\sethlcolor{yellow}

\newcommand{\biota}{\iota
                     \hskip-.15ex{\hbox to 0pt{\hss {\leavevmode
                     \hbox{\raise -.60ex \hbox{{\tt \'{}}}}}}}
                     \hskip.37ex{\hbox to 0pt{\hss {\leavevmode
                     \hbox{\raise -.50ex \hbox{{\tt \'{}}}}}}}}

\begin{document}

\preprint{00}

\title{
Radially local approximation of 
%
%
the
%
%
drift kinetic equation
}



\author{H. Sugama}
\affiliation{
National Institute for Fusion Science, 
Toki 509-5292, Japan
}
\affiliation{
Department of Fusion Science, SOKENDAI (The Graduate University for Advanced Studies), 
Toki 509-5292, Japan 
}

\author{S. Matsuoka}
\affiliation{
Japan Atomic Energy Agency, 178-4, Wakashiba, Kashiwa 277-0871, Japan
}

\author{S. Satake}
\affiliation{
National Institute for Fusion Science, 
Toki 509-5292, Japan
}
\affiliation{
Department of Fusion Science, SOKENDAI (The Graduate University for Advanced Studies), 
Toki 509-5292, Japan 
}

\author{R. Kanno}
\affiliation{
National Institute for Fusion Science, 
Toki 509-5292, Japan
}
\affiliation{
Department of Fusion Science, SOKENDAI (The Graduate University for Advanced Studies), 
Toki 509-5292, Japan 
}


\date{\today}

\begin{abstract}
   A novel radially local approximation 
of the drift kinetic equation is presented. 
   The new drift kinetic equation that includes both ${\bf E} \times {\bf B}$ 
and tangential magnetic drift 
terms is written in the conservative form and 
it has favorable properties for numerical simulation 
that any additional terms for particle and energy sources 
are unnecessary for 
obtaining stationary solutions under the radially local approximation. 
   These solutions satisfy the intrinsic ambipolarity condition for 
neoclassical particle fluxes in the presence of quasisymmetry
of the magnetic field strength. 
   Also, another radially local drift kinetic equation 
is presented, from which
the positive definiteness of entropy production due to
neoclassical transport and Onsager symmetry of neoclassical 
transport coefficients are derived 
while it sacrifices the ambipolarity condition for 
neoclassical particle fluxes in axisymmetric and quasi-symmetric systems.

\end{abstract}


\maketitle 



\section{INTRODUCTION}

    Effects of neoclassical transport~\cite{H&H,H&S,Helander} 
on plasma confinement are more significant in 
stellarator and heliotron plasmas
than in tokamak plasmas because, in the former, 
radial drift motions of trapped particles in helical ripples enhance particle and heat 
transport due to nonaxisymmetry of 
the magnetic configuration.~\cite{Wakatani,Spong,Shaing} 
    Conventional calculations of neoclassical transport fluxes are done applying  
radially local approximation to solving the drift kinetic equation, in which   
${\bf v}_d \cdot \nabla f$ are often neglected as a small term of 
higher order in the normalized gyroradius parameter $\delta \sim \rho / L$. 
(Here, ${\bf v}_d$, $f$, $\rho$, and $L$ represent the guiding center drift velocity,  
the deviation of the guiding center distribution function from the local Maxwellian 
equilibrium distribution, the gyroradius, and the equilibrium scale length, respectively.)
    However, in stellarator and heliotron plasmas, 
this ${\bf v}_d \cdot \nabla f$ term is known to  
be influential on the resultant neoclassical transport  because it significantly 
changes orbits of particles trapped in helical ripples. 
    Therefore, at least, 
the ${\bf E} \times {\bf B}$ drift part ${\bf v}_E \cdot \nabla f$ 
in ${\bf v}_d \cdot \nabla f$ 
has been kept  
in most studies of neoclassical transport 
in helical 
systems.~\cite{DKES,Beidler,Taguchi,Sugama2002,PENTA,Satake,Matsuoka,Landreman,Belli}

    Recently,  it was shown by Matsuoka {\it et al}.~\cite{Matsuoka} 
that the neoclassical transport is significantly 
influenced by retaining the magnetic drift tangential to flux surfaces 
in ${\bf v}_d \cdot \nabla f$ for the magnetic configuration of 
LHD especially when the radial electric field is weak. 
   However, as pointed by Landreman {\it et al}.,~\cite{Landreman}  
stationary solutions of the drift kinetic equation with 
radially local approximation used 
require additional artificial sources (or sinks) of particles and energy
when the above-mentioned drift terms are retained.  
    In this paper, a novel radially local drift kinetic equation, which  
 includes both ${\bf E} \times {\bf B}$ 
and tangential magnetic drift motions, is presented.   
  The radially local guiding center motion equations do not satisfy 
the conservation law of the phase-space volume while the full 
guiding center motion equations do.   
   This fact causes the difficulty in obtaining the stationary solution 
of the local drift kinetic equation. 
   However, the new local drift kinetic equation, which is 
written in the conservative form,  
has favorable properties for numerical simulation 
such that any additional terms for particle and energy sources 
are unnecessary for obtaining stationary solutions. 
   In addition, it satisfies the intrinsic ambipolarity condition for 
neoclassical particle fluxes in axisymmetric systems as well as 
in quasi-symmetric helical systems.~\cite{Helander2008,Sugama2011} 
   The present work also treats interesting issues 
regarding the entropy production rate and 
Onsager symmetry~\cite{deGroot,Sugama1996a} 
for neoclassical transport equations  
resulting from the new local drift kinetic model. 

The rest of this paper is organized as follows. 
   In Sec.~II, we consider the full drift kinetic model based 
on Littlejohn's guiding-center equations~\cite{Littlejohn} without
radially local approximation. 
  Particle, energy, and parallel momentum balance equations 
are derived from the full drift kinetic equation. 
   These balance equations are flux-surface averaged to 
confirm that they contain the second-order terms in $\delta$, 
which represent neoclassical transport across flux surfaces. 
   Also, expanding  the distribution function about the local Maxwellian, 
we rewrite the drift kinetic equation to explicitly show that the   
thermodynamic forces defined by the background density and
temperature gradients and the parallel electric field  
cause the deviation $f$ from the local Maxwellian.  
   In Sec.~II,  a new drift kinetic model is 
constructed by applying radially local approximation to  
Littlejohn's guiding-center equations with keeping 
${\bf E} \times {\bf B}$ 
and tangential magnetic drift velocities. 
   The new local drift kinetic equation for $f$ is shown  to 
be compatible with the stationary solution and 
to give 
intrinsic ambipolar particle fluxes for axisymmetric 
and quasi-symmetric systems. 
   In Sec.~IV,  we present another radially local drift kinetic equation,  
from which the positive definiteness of entropy production due to 
neoclassical transport and Onsager symmetry of neoclassical transport 
coefficients are derived although
this local drift kinetic equation no longer guarantees rigorously 
the intrinsic ambipolarity of neoclassical particle fluxes 
for axisymmetric and quasi-symmetric systems. 
  Finally, conclusions are given in Sec.~V.

\section{FULL DRIFT KINETIC MODEL}

\subsection{Drift kinetic model based on  
Littlejohn's guiding-center equations}

We denote the guiding-center variables by
$({\bf X}, U, \xi, \mu)$, where ${\bf X}$ represents 
the position vector of the 
guiding center, $U$ the parallel velocity, $\xi$ the gyrophase defined by 
the azimuthal angle of the gyroradius vector around the magnetic field line, 
and $\mu$ the magnetic moment. 
   The Lagrangian for the guiding-center motion is given by 
Littlejohn~\cite{Littlejohn} as
\begin{equation}
\label{Lagrangian}
L = \left( \frac{e}{c} {\bf A} + m U {\bf b} \right)
\cdot \dot{\bf X} + \frac{m c}{e} \mu \dot{\xi}
-H
, 
\end{equation}
where the Hamiltonian $H$ is given by
\begin{equation}
\label{Hamiltonian}
H =  \frac{1}{2} m U^2 + \mu B + e \Phi
. 
\end{equation}
Here, $\Phi$ denotes the electrostatic potential. 
   Using Eqs.~(\ref{Lagrangian}) and (\ref{Hamiltonian}), 
the guiding-center motion equations are derived as 
\begin{eqnarray}
\label{gu_eqs}
\frac{d {\bf X}}{dt}
& = & 
{\bf V}_{\rm gc}
\equiv 
U {\bf b} + \frac{c}{e B_\parallel^*}{\bf b}
\times
( m U^2 {\bf b}\cdot \nabla {\bf b} 
+ \mu \nabla B - e {\bf E}^* )
,
\nonumber \\
\frac{d U}{dt}
& = & 
-\frac{1}{m} {\bf b} \cdot  ( \mu \nabla B - e {\bf E} )
+ U  {\bf b}\cdot \nabla {\bf b} \cdot 
{\bf V}_{\rm gc}
,
\nonumber \\
\frac{d \xi}{dt}
& = & 
\Omega
,
\nonumber \\
\frac{d \mu}{dt}
& = & 
0
,
\end{eqnarray}
where $\Omega = e B/(mc)$, 
$\nabla = \partial / \partial {\bf X}$, 
${\bf E} \equiv  - \nabla \Phi - c^{-1} 
\partial {\bf A}/ \partial t$, 
${\bf B} \equiv \nabla \times {\bf A}$, 
${\bf E}^* \equiv  - \nabla \Phi - c^{-1} 
\partial {\bf A}^*/ \partial t$, 
${\bf B}^* \equiv \nabla \times {\bf A}^*$, 
$B_\parallel^* \equiv {\bf B}^*  \cdot {\bf b}$, 
and 
${\bf A}^* \equiv {\bf A} + (m c/e) U {\bf b}$ 
are used, and  
   the guiding-center drift velocity ${\bf V}_{\rm gc}$ is 
defined by the right-hand side of the equation 
for $d{\bf X}/dt$. 
   On the right-hand side of the equation for $dU/dt$ in 
Eq.~(\ref{gu_eqs}), 
the last term 
$U  {\bf b}\cdot \nabla {\bf b} \cdot 
{\bf V}_{\rm gc}$ 
is smaller than other terms by the order of 
$\delta = \rho /L$ where $\rho$ and $L$ represent 
the gyroradius and the gradient scale length given by 
$L \sim B/ |\nabla B | \sim \Phi / |\nabla \Phi |$. 

The Jacobian for the guiding-center variables is written as 
\begin{equation}
\label{D}
D
=
\det \left[ \frac{\partial ({\bf x}, {\bf v})}{\partial ({\bf X}, U, \xi, \mu)} \right]
=
\frac{B_\parallel^*}{m}
\end{equation}
where ${\bf x}$ and ${\bf v}$ denote the particle position vector and 
the velocity vector, respectively. 
    Then, the conservation of  the phase-space volume
$d^3 x \, d^3 v = D \, d^3 X \, dU \, d\xi \, d\mu$ is represented by 
\begin{equation}
\label{dDdt}
\frac{\partial  D }{\partial t}
+ 
\nabla \cdot  ( D \dot{\bf  X} )
+
\frac{\partial ( D  \dot{U} )}{\partial U}
=  0
, 
\end{equation}
which can be proved by using Eqs.~(\ref{gu_eqs}) and (\ref{D}).

The drift kinetic equation for the distribution function $F ( {\bf X}, U, \mu, t)$ 
is given by 
\begin{equation}
\label{DKE}
\left(
\frac{\partial }{\partial t}
+
\dot{\bf  X} \cdot \nabla
+ \dot{U}
\frac{\partial}{\partial U}
\right) 
F ( {\bf X}, U, \mu, t) =  C (F) + {\cal S}
\end{equation}
where the total time derivative is denoted by 
$\dot{} = d/dt$. 
In the right-hand side of Eq.~(\ref{DKE}), 
$C (F)$ is the collision term and the additional term ${\cal S}$ is 
given to represent external particle, momentum, and/or energy 
sources if any. 
   Here, ${\cal S}$ is considered to be of the second order in $\delta$. 
   We can also treat effects of turbulent fluctuations 
by Eq.~(\ref{DKE}) if we regard the second-order additional term 
${\cal S}$ as the ensemble average of 
the product of fluctuation parts in the electromagnetic fields and 
the distribution function as shown in 
Refs.~21 and 22 
where 
the notation ${\cal D}$ is used instead of ${\cal S}$ to represent 
the term including the effects of turbulent fluctuations. 
   Using Eq.~(\ref{dDdt}), the drift kinetic equation can be rewritten 
in the conservative form as 
\begin{equation}
\label{CDKE}
\frac{\partial  ( D F ) }{\partial t}
+ 
\nabla \cdot  ( D F \dot{\bf  X} )
+
\frac{\partial ( D F \dot{U})}{\partial U}
=  D  [ C (F) + {\cal S} ]
. 
\end{equation}

\subsection{Particle, energy, and parallel momentum balance equations}

     Multiplying Eq.~(\ref{CDKE}) with an arbitrary function 
${\cal A}(t, {\bf X}, U, \mu)$ which is independent of the gyrophase $\xi$ and 
taking its velocity-space integral, the balance equation for the 
density variable $\int d^3 v \, F {\cal A}$ in the ${\bf X}$-space 
is derived as 
\begin{eqnarray}
\label{consA}
& & \frac{\partial}{\partial t}
\left( \int d^3 v \, F {\cal A} \right) 
+ 
\nabla \cdot  
\left( \int d^3 v \, F {\cal A} \dot{\bf  X}  \right) 
\nonumber \\ 
& & 
=  
\int d^3 v \left( 
F \dot{\cal A} + [ C (F) + {\cal S} ] {\cal A}
\right) 
,
\end{eqnarray}
where 
\begin{equation}
\dot{\cal A} = \frac{d {\cal A} }{ dt }
= 
\frac{\partial {\cal A} }{\partial t }
+ \dot{\bf  X} \cdot \nabla {\cal A}
+ \dot{U} \frac{\partial {\cal A} }{\partial U}
, 
\end{equation}
and  the velocity-space integral is denoted by 
$\int d^3 v = 2 \pi    \int dU \int d\mu \, D $
 for gyrophase-independent integrands. 
   For the case of ${\cal A} = 1$, 
Eq.~(\ref{consA}) reduces to the time-evolution equation for the 
density  $\int d^3 v \, F$, 
\begin{equation}
\label{consn}
\frac{\partial}{\partial t}
\left( \int d^3 v \, F \right) 
+ 
\nabla \cdot  
\left( \int d^3 v \, F \dot{\bf  X}  \right) 
=
\int d^3 v \,
{\cal S} 
.
\end{equation}
   In deriving Eq.~(\ref{consn}), the 
conservation law, $ \int d^3 v \, C(F) = 0$, is used. 
   However, it is noted that, 
if we use the collision operator obtained by 
the transformation from the particle coordinates to 
the guiding-center coordinates with finite-gyroradius 
effects taken into account, 
the velocity-space integral $ \int d^3 v \, C(F)$ does not vanish but 
it becomes the opposite sign of the divergence of the classical particle flux 
as shown in Refs.~23--25. 
  Here and hereafter, we assume that the expression of $C(F)$ is 
the same as that of the Landau collision operator given in the particle 
coordinates for simplicity so that $\int d^3 v \, C(F) = 0$ is satisfied and 
the classical transport is neglected.   

   We next consider the energy 
$
{\cal E} 
=  
H
$ 
[see Eq.~(\ref{Hamiltonian})] 
as ${\cal A}$ in Eq.~(\ref{consA}) and 
obtain the energy balance equation, 
\begin{eqnarray}
& & 
\frac{\partial}{\partial t}
\left( \int d^3 v \, F  {\cal E}  \right) 
+ 
\nabla \cdot  
\left( \int d^3 v \, F  {\cal E}  \dot{\bf  X}  \right) 
\nonumber \\
& & =
\int d^3 v \,
\left( 
F \dot{\cal E} + [ C (F) + {\cal S} ] {\cal E}
\right) 
, 
\end{eqnarray}
   where the total time derivative of the energy is written as
\begin{eqnarray}
\label{dotE}
\dot{\cal E}
& = & 
\frac{d {\cal E} }{d t}
\nonumber \\
& = &  
 e \frac{\partial \Phi ({\bf X}, t) }{\partial t}
+ \mu \frac{\partial B ({\bf X}, t) }{\partial t}
- \frac{e}{c} \frac{\partial {\bf A}^* ({\bf X}, t) }{\partial t} \cdot 
\dot{\bf X}
. 
\end{eqnarray}
We easily see from Eq.~(\ref{dotE}) that $\dot{\cal E} = 0$ 
for the stationary electromagnetic field. 
 When we use the kinetic energy, 
\begin{equation}
W
=  
\frac{1}{2}m U^2 + \mu B
= {\cal E} - e \Phi
, 
\end{equation}
another form of the energy balance equation is given by 
\begin{eqnarray}
& & 
\frac{\partial}{\partial t}
\left( \int d^3 v \, F  W  \right) 
+ 
\nabla \cdot  
\left( \int d^3 v \, F  W \dot{\bf  X}  \right) 
\nonumber \\
& & =
\int d^3 v \,
\left( 
F \dot{W} + [ C (F) + {\cal S} ] W
\right) 
,
\end{eqnarray}
  where the total time derivative of the kinetic energy is written as
\begin{eqnarray}
\label{dotW}
\dot{W}
& = & 
\frac{d W}{d t}
= 
\frac{d {\cal E}}{d t} - e \frac{d \Phi}{d t}
\nonumber \\
& = &  
\mu \frac{\partial B ({\bf X}, t) }{\partial t}
+ e {\bf E}^* \cdot 
\dot{\bf X}
. 
\end{eqnarray}

The parallel momentum balance equation is derived from 
Eq.~(\ref{consA}) with ${\cal A} = m U$ as 
\begin{eqnarray}
\label{paramomb}
& & 
\frac{\partial}{\partial t}
\left( \int d^3 v \, F  m U \right) 
+ 
\nabla \cdot  
\left( \int d^3 v \, F  m U \dot{\bf  X}  \right) 
\nonumber \\
& & =
\int d^3 v \,
\left( 
F m \dot{U} + [ C (F) + {\cal S} ] m U
\right) 
. 
\end{eqnarray}
   We now use 
\begin{eqnarray}
& & 
\nabla \cdot  
\left( \int d^3 v \, F  m U \dot{\bf  X}  \right) 
- 
\int d^3 v \, F m \dot{U}
\nonumber \\
& = & 
\nabla \cdot   
\left( \int d^3 v \, F  m U^2 {\bf b} \right)
+
\int d^3 v \, F {\bf b} \cdot ( \mu \nabla B  - e {\bf E} )
\nonumber \\ &  & \mbox{}
+ 
\nabla \cdot   
\left( \int d^3 v \, F  m U \dot{\bf  X}_\perp \right)
- 
\int d^3 v \, F m U \dot{\bf  X}_\perp \cdot 
( {\bf b} \cdot \nabla ) {\bf b}
\nonumber \\
& = & 
{\bf b} \cdot 
\left( \nabla \cdot   
\left[ \int d^3 v \, F \left( 
m U^2 {\bf b} {\bf b}
+ \mu B ( {\bf I} - {\bf b} {\bf b} )
\right. \right. \right.
\nonumber \\
&  & 
\left. \left. \left.
\hspace*{5mm}
\mbox{} + 
m U ( \dot{\bf  X}_\perp {\bf b}
+ 
{\bf b} \dot{\bf  X}_\perp ) 
\right) \right] \right)
- e E_\parallel \int d^3 v \, F 
, 
\end{eqnarray}
  and rewrite Eq.~(\ref{paramomb}) as 
\begin{equation}
\label{paramomb2}
\frac{\partial}{\partial t}
( n m u_\parallel )
+ 
{\bf b} \cdot  
( \nabla \cdot  {\bf P} )
=
n e E_\parallel + F_\parallel 
+
\int d^3 v \,
{\cal S} m U
.
\end{equation}
   Here,  the density $n$, the parallel flow velocity $u_\parallel$, the 
pressure tensor ${\bf P}$, and the parallel friction force $F_\parallel$ 
are defined by
\begin{eqnarray}
\label{nupppf}
n 
& = & 
\int d^3 v \, F 
\nonumber \\
n u_\parallel
& = & 
\int d^3 v \, F U
 \nonumber \\
{\bf P}
& = & 
 {\bf P}_{\rm CGL}  + \mbox{\boldmath$\pi$}_2 
\nonumber \\
{\bf P}_{\rm CGL}  
& = & 
\int d^3 v \, F \left( 
m U^2 {\bf b} {\bf b}
+ \mu B ( {\bf I} - {\bf b} {\bf b} )
\right)
\nonumber \\
\mbox{\boldmath$\pi$}_2
& = & 
\int d^3 v \, F 
m U ( \dot{\bf  X}_\perp {\bf b}
+ 
{\bf b} \dot{\bf  X}_\perp ) 
\nonumber \\
F_\parallel
& = & 
\int d^3 v \,
 C (F) m U
, 
\end{eqnarray}
where $\dot{\bf  X}_\perp \equiv \dot{\bf  X} -  
(\dot{\bf  X} \cdot  {\bf b} ) {\bf b}$. 
   Note that the pressure tensor ${\bf P}$ consists of 
the Chew-Goldbeger-Low (CGL) tensor~\cite{H&M} ${\bf P}_{\rm CGL}$ 
and the viscosity tensor $\mbox{\boldmath$\pi$}_2$ of 
the second order in $\delta$, where 
$\mbox{\boldmath$\pi$}_2$ satisfies 
$\mbox{\boldmath$\pi$}_2 : {\bf I} = \mbox{\boldmath$\pi$}_2 : {\bf b}{\bf b} = 0$
and the deviation of $F$ from the local Maxwellian distribution is considered to be of 
${\cal O}(\delta)$. 
    
It is well known  that, if we use 
the original Boltzmann kinetic equation instead of the drift kinetic equation in 
Eq.~(\ref{CDKE}), we can derive the momentum balance equation, 
\begin{equation}
\label{MBE}
\frac{\partial}{\partial t}
( n m {\bf u})
+ 
\nabla \cdot  {\bf P} 
=
n e \left( 
{\bf E} + 
\frac{\bf u}{c}  \times {\bf B}
\right) 
+
{\bf F}
+
\int d^3 v \,
{\cal S} m {\bf v}
, 
\end{equation}
where the Boltzmann kinetic equation is assumed to also contain 
the source term ${\cal S}$. 
   In Eq.~(\ref{MBE}), the particle flow $n {\bf u}$, the pressure tensor ${\bf P}$, and 
the friction force ${\bf F}$ are defined by $n {\bf u} = \int d^3 v \, F$, 
${\bf P} = \int d^3 v \, F m {\bf v} {\bf v}$, and 
${\bf F}  = \int d^3 v \, C(F) m {\bf v} $, where, exactly speaking,   
$F=F({\bf x}, {\bf v}, t)$ represents the particle distribution function given by the solution 
of the Boltzmann kinetic equation and it has a gyrophase dependence that is not 
included in the solution of the drift kinetic equation.  
   Comparing Eqs.~(\ref{paramomb2}) and (\ref{MBE}), we see that 
Eq.~(\ref{paramomb2}) coincides with the parallel component 
of the exact momentum balance equation in Eq.~(\ref{MBE}) 
except that the former contains 
$nm {\bf u} \cdot \partial {\bf b}/\partial t$ and the non-CGL viscosity tensor 
expressed differently from the one in the latter. 

We now consider general toroidal configurations, for which the magnetic field 
is written in terms of the flux coordinates $(s, \theta, \zeta)$ as 
\begin{equation}
\label{B}
{\bf B} 
=
\psi'  \nabla s \times \nabla \theta +
\chi' \nabla \zeta \times \nabla s
, 
\end{equation}
where $\theta$ and $\zeta$ represent the poloidal and toroidal angles,
respectively, and $s$ is an arbitrary label of a flux surface. 
The poloidal and toroidal fluxes within a flux surface labeled by $s$ 
are given by $2 \pi \psi (s)$ and $2 \pi \chi (s)$􏰩􏰦, respectively. 
The derivative with respect to $s$ is denoted by 􏱆􏰳$' = d/ds$ so that 􏰘􏱆􏰳
$\psi'  = d􏰘\psi/ds$ and 􏰥􏱆􏰳$\chi'  = d􏰘\chi/ds$. 
%
%
  Taking the flux-surface average of the covariant toroidal component of Eq.~(\ref{MBE}) 
and making the summation over species,
%
%
we obtain the expression for the radial current as~\cite{Shaing} 
\begin{eqnarray}
\label{rcurrent0}
\frac{\chi' }{c} \sum_a e_a 
\langle n_a u_a^s \rangle
& = &
\sum_a 
\left[
m_a
\frac{\partial}{\partial t}
\langle n_a u_{a\zeta} \rangle
+ 
 \langle 
( \nabla \cdot  {\bf P}_a )_\zeta
\rangle
\right. 
\nonumber \\
& & 
\left. \mbox{}
-
 \left\langle 
\int d^3 v \,
%
%
{\cal S}_a 
%
%
m_a v_\zeta
\right\rangle
\right]
, 
\end{eqnarray}
where the superscript $s$ and 
the subscript $\zeta$ represent the covariant radial component and 
contravariant toroidal component  given by taking 
the inner products with $\nabla s$ and $\partial {\bf x}/\partial \zeta$,
respectively, and 
the subscript $a$ is used to explicitly show the particle species. 
   Using the symmetry property of the pressure tensor ${\bf P}$, 
we can show that, for axisymmetric toroidal systems, 
\begin{equation}
\label{divPz}
 \langle 
( \nabla \cdot  {\bf P} )_\zeta
\rangle
= 
\frac{1}{V'}
\frac{\partial }{\partial s}
( V' \langle P_\zeta^s \rangle )
,
\end{equation}
where 
$P_\zeta^s = \nabla s \cdot {\bf P} \cdot \partial {\bf x}/\partial \zeta$. 
In axisymmetric and 
quasi-axisymmetric toroidal systems,~\cite{Sugama2011}
  we have 
\begin{equation}
\label{divPCGLz}
 \langle 
( \nabla \cdot  {\bf P}_{\rm CGL} )_\zeta
\rangle
= 
0
.
\end{equation}
   Then, using Eqs.~(\ref{rcurrent0})--(\ref{divPCGLz}) and 
${\bf P} =
 {\bf P}_{\rm CGL}  + \mbox{\boldmath$\pi$}_2$, 
we find that,  
 even for axisymmetric toroidal systems in the stationary state 
 ($\partial /\partial t = 0$) with ${\cal S}=0$, 
the surface-averaged radial current does not vanish exactly due to 
the second-order viscosity tensor $\mbox{\boldmath$\pi$}_2$ as shown by 
\begin{eqnarray}
\label{pa2sz}
\frac{\chi' }{c} \sum_a e_a 
\langle n_a u_a^s \rangle
& = &
\sum_a 
\frac{1}{V'}
\frac{\partial }{\partial s}
[ V' \langle (\pi_{a2})_\zeta^s \rangle ]
. 
\end{eqnarray}
   However, it is shown in Ref.~17 
that 
$\langle (\pi_{a2})_\zeta^s \rangle$ is a small quantity of ${\cal O}(\delta^3)$
in axisymmetric systems with up-down symmetry (as well as in quasi-axisymmetric 
systems with stellarator symmetry) where all terms in  
the toroidal momentum balance equation given from Eq.~(\ref{rcurrent0}) 
vanish up to ${\cal O}(\delta^2)$. 
   The same argument as above can be done for other quasi-symmetric systems such as 
quasi-poloidally-symmetric and quasi-helically-symmetric 
systems if stellarator symmetry holds. 
%
%
On the other hand, 
in axisymmetric systems without up-down symmetry, 
$\langle (\pi_{a2})_\zeta^s \rangle = {\cal O}(\delta^3)$ 
is not guaranteed. 
Then, the ambipolarity condition 
$
\sum_a e_a 
\langle n_a u_a^s \rangle
= 0
$
is not automatically satisfied on the second order in  $\delta$ 
because of the third-order radial particle fluxes 
$(c/e_a \chi' V') 
\partial [ V' \langle (\pi_{a2})_\zeta^s \rangle ]/\partial s$
driven by 
the second-order shear viscosity tensor components $(\pi_{a2})_\zeta^s$
[here, it is useful to formally regard 
the electric charge as the ${\cal O}(\delta^{-1})$ 
quantity~\cite{Littlejohn1981}
so that the radial current due to the third-order radial particle flux 
is immediately found to be of the second order]. 
However, even in this axisymmetric 
but up-down asymmetric case, the second-order radial 
neoclassical particle fluxes driven by the CGL tensors 
still automatically satisfy the ambipolarity condition for the radial current 
up to the first order.~\cite{H&H,H&S,Helander}
%
%

\subsection{Drift kinetic equation expressed in terms of flux coordinates}

Using the flux coordinates $(s, \theta, \zeta)$, 
the drift kinetic equation, Eq.~(\ref{DKE}), is rewritten as
\begin{eqnarray}
\label{DKE1}
& & 
\left(
\frac{\partial }{\partial t}
+ \dot{s}
\frac{\partial}{\partial s}
+ \dot{\theta}
\frac{\partial}{\partial \theta}
+ \dot{\zeta}
\frac{\partial}{\partial \zeta}
+ \dot{U}
\frac{\partial}{\partial U}
\right) 
F ( s, \theta, \zeta, U, \mu, t) 
\nonumber \\
& & =  C (F) + {\cal S}
\end{eqnarray}
   where 
\begin{eqnarray}
\label{dotstz}
& & 
[\dot{s} , \dot{\theta}, \dot{\zeta} ] 
=
\frac{d}{dt} [s ,  \theta ,  \zeta ] 
\nonumber \\ 
& & =
\left(
\frac{\partial}{\partial t} + \dot{\bf X} \cdot \nabla
\right)
[ s({\bf X}, t), \theta({\bf X}, t),  \zeta({\bf X}, t) ]
.
\end{eqnarray}
   In Eq.~(\ref{dotstz}),  the functions 
$s({\bf X}, t)$, $\theta({\bf X}, t)$,  and $\zeta({\bf X}, t)$ are 
defined by the inverse of 
${\bf X} = {\bf X}(s, \theta, \zeta, t)$, 
where $t$ is generally included as a parameter. 
   Denoting the Jacobian for the flux coordinates $(s ,  \theta ,  \zeta)$ 
by 
\begin{equation}
\sqrt{g} 
=
\det \left[ \frac{\partial ({\bf X})}{\partial (s, \theta, \zeta)} \right]
=
\frac{1}{
[\nabla s \cdot (\nabla \theta \times \nabla \zeta) ]
}
, 
\end{equation}
the conservation law of the phase-space volume, Eq.~(\ref{dDdt}),  and 
the conservative form of the drift kinetic equation, Eq.~(\ref{CDKE}), 
are rewritten as  
\begin{eqnarray}
& & 
\frac{\partial  ( \sqrt{g} D ) }{\partial t}
+ 
\frac{\partial ( \sqrt{g} D \dot{s} )}{\partial s}
+ 
\frac{\partial ( \sqrt{g} D  \dot{\theta} )}{\partial \theta}
+ 
\frac{\partial ( \sqrt{g} D  \dot{\zeta} )}{\partial \zeta}
\nonumber \\
& & 
\mbox{}
+
\frac{\partial ( \sqrt{g} D  \dot{U} )}{\partial U}
=  0, 
\end{eqnarray}
   and 
\begin{eqnarray}
\label{gDFeq}
& & 
\frac{\partial  ( \sqrt{g} D F ) }{\partial t}
+ 
\frac{\partial ( \sqrt{g} D F\dot{s} )}{\partial s}
+ 
\frac{\partial ( \sqrt{g} D F \dot{\theta} )}{\partial \theta}
+ 
\frac{\partial ( \sqrt{g} D F \dot{\zeta} )}{\partial \zeta}
\nonumber \\
& & 
\mbox{}
+
\frac{\partial ( \sqrt{g} D F \dot{U} )}{\partial U}
=  
 \sqrt{g}  D [ C (F) + {\cal S} ] 
, 
\end{eqnarray}
respectively. 

For an arbitrary function 
${\cal A} (s, \theta, \zeta, U, \mu, t)$ 
which is independent of the gyrophase $\xi$, 
the phase-space integral is written as 
\begin{eqnarray}
& & 2 \pi \int d^3 X  \int  dU \int  d\mu 
\, D \, {\cal A} 
\nonumber \\
&  &
=
2 \pi \int ds \oint d\theta \oint d\zeta \sqrt{g} 
   \int dU \int d\mu \, 
D  \, {\cal A} 
\nonumber \\
& & 
=
\int ds \,
V'  
\left\langle \int d^3 v \,  {\cal A} \right\rangle
,
\end{eqnarray}
   where 
\begin{equation}
\label{saverage}
\langle \cdots \rangle
=
\frac{1}{V'}
\oint d\theta \oint d\zeta \sqrt{g} \cdots
\end{equation}
represents the flux-surface average and
\begin{equation}
V' = \frac{dV}{ds} 
=  
\oint d\theta \oint d\zeta \sqrt{g} 
\end{equation}
denotes the radial derivative of the volume $V(s)$ enclosed within a flux surface 
labeled by $s$. 
   We now integrate Eq.~(\ref{gDFeq}) with respect to  
the coordinates $(\theta, \zeta, U, \mu)$ to obtain
\begin{eqnarray}
\label{dFAdt}
& & 
\frac{\partial }{\partial t}
\left( 
V' 
\left\langle
\int d^3 v \; F {\cal A}
\right\rangle
\right)
+ 
\frac{\partial }{\partial s}
\left( 
V' 
\left\langle
\int d^3 v \; F {\cal A} \, \dot{s}
\right\rangle
\right)
\nonumber \\ 
& & 
=  
V' 
\left\langle
\int d^3 v \left( 
F \dot{\cal A}  + [ C (F) + {\cal S} ]  {\cal A}
\right) 
\right\rangle, 
\end{eqnarray}
 where 
\begin{equation}
\dot{\cal A} = \frac{d {\cal A} }{ dt }
= 
\frac{\partial {\cal A} }{\partial t }
+ \dot{s} \frac{\partial {\cal A} }{\partial s}
+ \dot{\theta} \frac{\partial {\cal A} }{\partial \theta}
+ \dot{\zeta} \frac{\partial {\cal A}}{\partial \zeta}
+ \dot{U} \frac{\partial {\cal A} }{\partial U}
. 
\end{equation}
   The time-evolution equation for the surface-averaged 
density $\left\langle \int d^3 v \, F \right\rangle$
is derived from Eq.~(\ref{dFAdt}) with ${\cal A}=1$ 
as 
\begin{eqnarray}
\label{s-particle-balance}
& & 
\frac{\partial }{\partial t}
\left( 
V' 
\left\langle
\int d^3 v \; F 
\right\rangle
\right)
+ 
\frac{\partial }{\partial s}
\left( 
V' 
\left\langle
\int d^3 v \; F  \, \dot{s}
\right\rangle
\right)
\nonumber \\ 
& & 
=  
V' 
\left\langle
\int d^3 v \, {\cal S} 
\right\rangle
.
\end{eqnarray}
   For the cases of 
${\cal A} = W$, 
Eq.~(\ref{dFAdt}) reduces to the 
surface-averaged energy balance equation, 
\begin{eqnarray}
\label{s-energy-balance}
& & 
\frac{\partial }{\partial t}
\left( 
V' 
\left\langle
\int d^3 v \; F W 
\right\rangle
\right)
+ 
\frac{\partial }{\partial s}
\left( 
V' 
\left\langle
\int d^3 v \; F W  \, \dot{s}
\right\rangle
\right)
\nonumber \\ 
& & 
=  
V' 
\left\langle
\int d^3 v \left(
F \dot{W}  + 
[ C (F) + {\cal S} ]  W
 \right)
\right\rangle
,
\end{eqnarray}
where $\dot{W}$ is given by Eq.~(\ref{dotW}). 
   In Eqs.~(\ref{s-particle-balance}) and (\ref{s-energy-balance}), 
$\left\langle \int d^3 v \; F \, \dot{s} \right\rangle$ and 
$\left\langle \int d^3 v \; F W  \, \dot{s} \right\rangle$
represent 
the radial neoclassical transport fluxes of particles and energy, respectively, 
which are regarded as of ${\cal O}(\delta^2)$ 
assuming that the deviation of $F$ from the local Maxwellian is of 
${\cal O}(\delta)$ (see Sec.~II.D).  
  The radial transport fluxes of ${\cal O}(\delta^2)$ are consistent with 
the so-called transport ordering~\cite{H&H} which implies 
$\partial / \partial t = {\cal O}(\delta^2)$ in 
Eqs.~(\ref{s-particle-balance}) and (\ref{s-energy-balance}).

\subsection{Expansion about a local Maxwellian distribution}

The zeroth-order solution $F_0$ of the drift kinetic equation, 
Eq.~(\ref{DKE1}), is given by the local Maxwellian, 
\begin{eqnarray}
\label{Maxwellian}
F_0
& = & 
n_0 
\left(
\frac{m}{2\pi T_0}
\right)^{3/2}
\exp 
\left( - \frac{W}{T_0} \right)
\nonumber \\ 
& = & 
n_0 
\left(
\frac{m}{2\pi T_0}
\right)^{3/2}
\exp 
\left( - \frac{{\cal E} - e \Phi}{T_0} \right)
, 
\end{eqnarray}
which annihilates the collision term,
\begin{equation}
 C (F_0) = 0
. 
\end{equation}
    The total time derivative of $F_0$ is written as 
\begin{eqnarray}
\frac{d F_0}{d t}  & = & 
F_0 
\left[
\frac{d \ln n_0}{dt} 
+ \frac{d \ln T_0}{dt} 
\left( \frac{m v^2}{2 T_0} - \frac{3}{2} \right)
- \frac{1}{T_0} \frac{d W}{dt} 
\right]
\nonumber \\ 
& = & 
F_0 
\left\{
\dot{\bf X}\cdot \nabla s
\left[ 
\frac{\partial \ln n_0}{\partial s} + 
\frac{e}{T_0} \frac{\partial \langle \Phi \rangle}{\partial s}
+ \frac{\partial \ln T_0}{\partial s} 
\right. \right.
\nonumber \\ & & 
\mbox{} \left. \left. 
\times
\left( \frac{W}{T_0} - \frac{3}{2} \right)
\right]
- \frac{e UE_\parallel }{T_0} 
\right\}
+ {\cal O}(\delta^2)
, 
\end{eqnarray}
   where the zeroth-order density $n_0$ and temperature $T_0$ are 
flux-surface functions independent of $(\theta, \zeta)$, and 
their time dependence follows the transport ordering, 
$\partial / \partial t = {\cal O}(\delta^2)$. 
    The parallel electric field $E_\parallel$ is given by
\begin{eqnarray}
\label{Epara}
E_\parallel
&  = &
- {\bf b} \cdot 
\left( \nabla \widetilde{\Phi} + \frac{1}{c}
\frac{\partial {\bf A}}{\partial t}
\right)
\nonumber \\ 
& = & 
B
\frac{\langle B E_\parallel \rangle }{\langle B^2 \rangle} 
+ \left( 
E_\parallel - B
\frac{\langle B E_\parallel \rangle }{\langle B^2 \rangle} 
\right)
,
\end{eqnarray}
    where $\widetilde{\Phi}= \Phi - \langle \Phi \rangle$. 
 We now define the first-order distribution $f$ by 
\begin{eqnarray}
\label{deff}
F
&  = &
F_0 \left[ 1 +  \frac{e}{T_0} \int^l
dl 
\left( 
E_\parallel - B
\frac{\langle B E_\parallel \rangle }{\langle B^2 \rangle} 
\right)
\right] 
+ f
, 
\end{eqnarray}
   where $\int^l dl$ represents the integral along the magnetic field line.  
   Then, substituting Eqs.~(\ref{Epara}) and (\ref{deff}) into 
Eq.~(\ref{DKE1}) yields
\begin{eqnarray}
\label{DKEf}
\frac{df}{dt}
&  = &
\frac{F_0}{T_0}
\left\{
V_{\rm gc}^s 
\left[  X_1  +  X_2 
\left( \frac{W}{T_0} - \frac{5}{2} \right)
\right]
+ \frac{e  UB}{\langle B^2 \rangle^{1/2}} X_E
\right\}
\nonumber \\ & & 
\mbox{}
+ C^L (f)
+ {\cal O}(\delta^2)
, 
\end{eqnarray}
   where the thermodynamic forces are defined by 
\begin{equation}
X_1
=  
- \frac{1}{n_0} \frac{\partial p_0}{\partial s}
-  e \frac{\partial \Phi}{\partial s}
, 
\hspace*{5mm}
X_2
=  
- \frac{\partial T_0}{\partial s}
, 
\hspace*{5mm}
X_E
=  
\frac{\langle B E_\parallel \rangle}{
\langle B^2 \rangle^{1/2}}
, 
\end{equation}
  and $C^L(f)$ represents the linearized collision operator. 
Note that all terms explicitly shown on the right-hand side of 
Eq.~(\ref{DKEf}) are
of the first order in $\delta$. 
   Using the transport ordering 
$\partial / \partial t = {\cal O}(\delta^2)$ and
$f = {\cal O}(\delta)$, 
the left-hand side of Eq.~(\ref{DKEf}) is written as 
\begin{eqnarray}
\label{dfdt2}
\frac{df}{dt}
&  = &
\left(
V_{\rm gc}^s 
\frac{\partial}{\partial s}
+ V_{\rm gc}^\theta
\frac{\partial}{\partial \theta}
+ V_{\rm gc}^\zeta
\frac{\partial}{\partial \zeta}
+ \dot{U}
\frac{\partial}{\partial U}
\right) 
f ( s, \theta, \zeta, U, \mu) 
\nonumber \\ & & 
\mbox{}
+ {\cal O}(\delta^3)
\nonumber \\
& = & 
\frac{1}{\cal D} 
\left(
\frac{\partial ( {\cal D} f V_{\rm gc}^s  )}{\partial s}
+ 
\frac{\partial (  {\cal D} f   V_{\rm gc}^\theta  )}{\partial \theta}
+ 
\frac{\partial (  {\cal D} f  V_{\rm gc}^\zeta  )}{\partial \zeta}
+
\frac{\partial ( {\cal D} f  \dot{U} )}{\partial U}
\right) 
\nonumber \\ & & 
\mbox{}
+ {\cal O}(\delta^3)
, 
\end{eqnarray}
where 
$V_{\rm gc}^s  = \dot{\bf X} \cdot \nabla s$, 
$V_{\rm gc}^\theta  = \dot{\bf X} \cdot \nabla \theta$, 
$V_{\rm gc}^\zeta  = \dot{\bf X} \cdot \nabla \zeta$, 
and 
${\cal D} =  \sqrt{g} D$. 
Since $V_{\rm gc}^s  =  {\cal O}(\delta)$, 
the radial drift term $V_{\rm gc}^s \partial f/\partial s$ 
in Eq.~(\ref{dfdt2})
is of the second order in $\delta$ and 
this gives rise to global or finite-orbit-width effects 
on neoclassical transport.

\section{RADIALLY LOCAL APPROXIMATION}

Under the radially local approximation made here, 
the guiding center equations are written as 
\begin{eqnarray}
\label{gu_eqs_rl}
\frac{d {\bf X}}{dt}
& = & {\bf V}_{\rm gc}^{\rm (rl)}
\equiv
U {\bf b} + ({\bf V}_{\rm gc}^{\rm (rl)} )_\perp
,
\nonumber \\
\frac{d U}{dt}
& = & 
-\frac{\mu }{m} {\bf b} \cdot  \nabla B 
+ U  {\bf b}\cdot \nabla {\bf b} \cdot 
{\bf V}_{\rm gc}^{\rm (rl)}
\nonumber \\
\frac{d \mu}{dt}
&  = & 
-
\frac{1}{B}  ({\bf V}_{\rm gc}^{\rm (rl)})_\perp
\cdot ( m U^2 {\bf b} \cdot \nabla {\bf b} + \mu \nabla B )
, 
\end{eqnarray}
   where the second-order part 
$-\nabla \widetilde{\Phi} - c^{-1} \partial {\bf A}/\partial t$ of 
the electric field ${\bf E}$  
is neglected and 
the guiding center drift velocity in Eq.~(\ref{gu_eqs}) is replaced by 
${\bf V}_{\rm gc}^{\rm (rl)}$  
 which has no radial component:  
${\bf V}_{\rm gc}^{\rm (rl)} \cdot \nabla s = 0$. 
   The component of ${\bf V}_{\rm gc}^{\rm (rl)}$ 
perpendicular to the magnetic field is denoted by
by $({\bf V}_{\rm gc}^{\rm (rl)} )_\perp$. 
  We later impose the condition, 
$({\bf V}_{\rm gc}^{\rm (rl)} )_\perp (\mu = 0 ) = 0$, 
in order to derive appropriate balance equations of 
particles, energy, and parallel momentum [see 
Eqs.~(\ref{continuity_eq}), (\ref{energy_balance}), 
and (\ref{para_mbalance})] by removing  
improper sources and/or sinks at the 
boundary $\mu = 0$ in the velocity-space integral domain.

   In Eq.~(\ref{gu_eqs_rl}), the magnetic 
moment $\mu$ is allowed to vary in time such that 
conservation of the kinetic energy of the particle 
$W =  m U^2/2 + \mu B$, 
\begin{equation}
\frac{d W}{dt}
 =  
m U \frac{d U}{dt} + B \frac{d\mu}{dt}
+
\mu {\bf V}_{\rm gc}^{\rm (rl)} \cdot \nabla B 
 = 0
, 
\end{equation}
is satisfied. 
   It might appear that 
the energy ${\cal E} = W + e \Phi$ should be conserved instead of $W$. 
   However, using $\Phi \simeq \langle \Phi \rangle$, 
we find that the difference $e \Phi$ between ${\cal E}$ and $W$ 
is approximately constant along the radially local 
guiding center orbit and accordingly the conservation of $W$ is reasonable 
under the radially local approximation. 

We now define $({\bf V}_{\rm gc}^{\rm (rl)})_\perp$ by 
removing the radial component from $({\bf V}_{\rm gc})_\perp$ 
as 
\begin{eqnarray}
\label{Vperp}
({\bf V}_{\rm gc}^{\rm (rl)})_\perp
&  =  &
\alpha (\Lambda)
\left(
({\bf V}_{\rm gc})_\perp  - 
\frac{({\bf V}_{\rm gc})_\perp \cdot \nabla s}{|\nabla s|^2} \nabla s
\right)
\nonumber \\ 
& = & 
\alpha (\Lambda)
\frac{c}{eB}
({\bf b}\times \nabla s)
\left(
\frac{\nabla s}{| \nabla s|^2} \cdot
[m v_\parallel^2 {\bf b} \cdot \nabla {\bf b} + \mu \nabla B ]
\right.
\nonumber \\ &  & 
\left. \mbox{}
+ e \frac{d\Phi}{ds}
\right)
\nonumber \\ & = & 
\alpha (\Lambda)
\frac{c}{eB}
({\bf b}\times \nabla s)
\left[
\left( \frac{m 
%
%
U^2
%
%
}{B} + \mu  \right)
\frac{\nabla s}{| \nabla s|^2} \cdot \nabla B
\right.
\nonumber \\ &  & 
\left. \mbox{}
+ m 
%
%
U^2 
%
%
\frac{4\pi}{B^2}\frac{d P}{ds}
+ e \frac{d\Phi}{ds}
\right]
,
\end{eqnarray}
where  
%
%
$B_\parallel^*$ and ${\bf E}^*$ in the definition of ${\bf V}_{\rm gc}$ given 
by Eq.~(\ref{gu_eqs}) are replaced with their lowest-order parts $B$ and 
$- (d\Phi/ds)\nabla s$, respectively, and 
%
%
the factor $\alpha (\Lambda)$ is introduced to 
satisfy the condition $({\bf V}_{\rm gc}^{\rm (rl)} )_\perp (\mu = 0 ) = 0$. 
 Here, the ratio of the magnetic moment $\mu$ to  the kinetic energy 
$W= m U^2/2 + \mu B$ is used to define the dimensionless parameter, 
$
\Lambda \equiv \mu B_{\rm max} / W
$, 
where $B_{\rm max}$ is the maximum value of $B$ on the flux surface. 
%
%
This parameter $\Lambda$ is a measure for classifying the guiding center motion  
into either passing or trapped orbit. 
As $\Lambda$ increases from 0 and approaches to 1, 
the orbit changes from the passing to the trapped one.  
%
%
   Then, we assume that 
\begin{equation}
\lim_{\Lambda \rightarrow +0}\alpha (\Lambda)
 =  0
\end{equation}
while $\alpha (\Lambda) = 1$ except for an interval,  
$0 \leq \Lambda < \Lambda_0$, where 
$\Lambda_0 ( \ll 1)$ is a small positive constant value. 
  For example, 
$\alpha (\Lambda)$
is defined by
\begin{equation}
\alpha (\Lambda)
 =  
\left\{
\begin{array}{lr}
\sin (\pi \Lambda/2\Lambda_0 )
& (\Lambda < \Lambda_0)
\\
1
& (\Lambda \geq \Lambda_0)
.
\end{array}
\right.
\end{equation}

  We should note that 
influences of the magnetic and ${\bf E}\times {\bf B}$ drift motions 
are significant mainly for precession drift orbits of trapped particles, 
and that particles in the region, $\Lambda < \Lambda_0$, 
are passing ones whose orbits almost coincide with field lines.  
   Therefore,  even if the functional form of $\alpha (\Lambda)$ and the value 
of $\Lambda_0$ are changed, 
the artificial reduction factor $\alpha (\Lambda)$ for $\Lambda < \Lambda_0$ 
is expected to cause little change in resultant passing particles' orbits except that 
the limiting condition, 
$\lim_{\Lambda \rightarrow +0} ({\bf V}_{\rm gc}^{\rm (rl)} )_\perp = 0$,  
is rigorously satisfied. 
   However, this insensitivity to the form of $\alpha (\Lambda)$ 
remains a future subject to be verified by numerical simulations. 

%
%
It also should be mentioned that the radially local approximation described by 
Eqs.~(\ref{gu_eqs_rl}) and (\ref{Vperp}) is independent of what 
poloidal and toroidal angles are chosen for the flux coordinates. 
This is a favorable property that is lost in Ref.~13. 
%
%
We see that the radially local guiding center equations given by  
Eqs.~(\ref{gu_eqs_rl}) and the Jacobian $D = B_\parallel^* / m$  for the phase-space coordinates $({\bf X}, U, \xi, \mu)$ [see Eq.~(\ref{D})] 
do not satisfy the conservation law of the phase-space volume as shown in 
Eq.~(\ref{dDdt}). 
%
%
This violation of the phase-space-volume conservation occurs even if $B_\parallel^*$ is used 
instead of $B$ in the denominator on the right-hand side of Eq.~(\ref{Vperp}) 
to define $({\bf V}_{\rm gc}^{\rm (rl)})_\perp$.
%
%
In the next section, we consider another Jacobian in order to recover 
the conservation law although,  in this section, 
a simpler approximate Jacobian $D_0 \equiv B/m$ is used. 
   Also, we hereafter employ $({\bf X}, W, U, \xi)$ as phase-space coordinates. 
   Then,  from the Jacobian $D_0 \equiv B/m$ for $({\bf X}, U, \xi, \mu)$
 with $\mu = (W - \frac{1}{2}m U^2)/B$, 
the Jacobian for $({\bf X}, W, U, \xi)$ is derived as $1/m$, 
which is constant in the phase space. 
      
Using ${\bf V}_{\rm gc}^{\rm (rl)}$, $dU/dt$, and 
$dW/dt = 0$ given by Eqs.~(\ref{gu_eqs_rl}) with (\ref{Vperp}) 
under the radially local approximation, 
the drift kinetic equation for the first-order distribution 
function $f({\bf X}, W, U)$ in the stationary state is written as 
\begin{eqnarray}
\label{dke_rl}
& & 
\nabla \cdot
(  f   {\bf V}_{\rm gc}^{\rm (rl)}   )
+
\frac{\partial}{\partial U}
\left(
 f  
\frac{dU}{dt}
\right)
\nonumber \\
&  = &
\frac{F_0}{T_0}
\left\{
V_{\rm gc}^s
\left[  X_1  +  X_2 
\left( \frac{W}{T_0} - \frac{5}{2} \right)
\right]
+ \frac{e U B}{\langle B^2 \rangle^{1/2}} X_E
\right\}
\nonumber \\ & & 
\mbox{}
+ C^L (f)
.  
\end{eqnarray}
  The radial component of the guiding center drift velocity 
$V_{\rm gc}^s$ 
on the right-hand side of Eq.~(\ref{dke_rl}) is given by
\begin{equation}
\label{vgcs}
V_{\rm gc}^s
= 
\frac{c}{eB^2}
[\nabla{s}\cdot 
({\bf b} \times \nabla B) ]
\left( 
\frac{1}{2}m U^2 + W 
\right)
. 
\end{equation}
In deriving  Eq.~(\ref{vgcs}) from the guiding center drift velocity 
given in Eq.~(\ref{gu_eqs}), 
only the lowest-order terms in $\delta$ is retained and 
the formula, 
$\nabla s \cdot [{\bf b} \times ({\bf b} \cdot \nabla) {\bf b} ] 
= \nabla s \cdot ({\bf b} \times \nabla B) /B$,  
obtained from the MHD equilibrium condition 
$\nabla [n_0(s)T_0(s)] = (4\pi)^{-1}(\nabla \times {\bf B}) \times {\bf B}$
is used. 
   The fact that the Jacobian is constant is used in deriving Eq.~(\ref{dke_rl})  
   which is rewritten by using 
the flux surface coordinates $(s, \theta, \zeta)$ as 
\begin{eqnarray}
\label{dke_rl2}
& &
\frac{1}{\sqrt{g}} 
\left[
\frac{\partial }{\partial \theta}
(  \sqrt{g} f   {\bf V}_{\rm gc}^{\rm (rl)} \cdot \nabla \theta  )
+ 
\frac{\partial }{\partial \zeta}
(  \sqrt{g}  f   {\bf V}_{\rm gc}^{\rm (rl)} \cdot \nabla \zeta  )
\right]
\nonumber \\ 
&  & 
\mbox{}
+
\frac{\partial}{\partial U}
\left(
 f  
\frac{dU}{dt}
\right)
\nonumber \\ 
& = & 
\frac{F_0}{T_0}
\left\{
V_{\rm gc}^s
\left[  X_1  +  X_2 
\left( \frac{W}{T_0} - \frac{5}{2} \right)
\right]
+ \frac{e U B}{\langle B^2 \rangle^{1/2}} X_E
\right\}
\nonumber \\ & & 
\mbox{}
+ C^L (f)
.
\end{eqnarray}
    Here, we should note that, in Eq.~(\ref{dke_rl2}),  
partial derivatives of the first-order distribution 
function $f$ are taken only with respect to 
the three variables $(\theta, \zeta, U)$ 
and that 
the radial coordinate $s$ and 
the kinetic energy $W$ enter $f(s, \theta, \zeta, W, U)$ as constant parameters.

   Taking the velocity-space integral of Eq.~(\ref{dke_rl}) yields 
the continuity equation in the stationary state, 
\begin{equation}
\label{continuity_eq}
\nabla \cdot 
( \Gamma_\parallel {\bf b} 
+ 
\mbox{\boldmath$\Gamma$}_{\perp  1} 
+ 
\mbox{\boldmath$\Gamma$}_{\perp  2}^{\rm (rl)} ) 
=  
0
.
\end{equation}
   The parallel  and perpendicular particle fluxes in Eq.~(\ref{continuity_eq}) 
are defined by 
\begin{eqnarray}
\Gamma_\parallel 
&  = &
n_0 u_\parallel
=
\int d^3 v \, f U
,
\nonumber \\
\mbox{\boldmath$\Gamma$}_{\perp  1} 
&  = &
n_0 {\bf u}_{\perp 1}
= 
\frac{n_0 c X_1}{e B} 
\nabla s \times  {\bf b}
 ,
\nonumber \\
\mbox{\boldmath$\Gamma$}_{\perp  2}^{\rm (rl)} 
&  = &
n_0 {\bf u}_{\perp 2}^{\rm (rl)} 
=
\int d^3 v \, f ({\bf V}_{\rm gc}^{\rm (rl)})_\perp
, 
\end{eqnarray}
where the velocity-space integral is written in terms of 
the variables $W$ and $U$ by 
\begin{equation}
\label{intv}
\int d^3 v = \frac{2 \pi}{m} 
\int_0^{+\infty} d W 
\int_{-\sqrt{2W/m}}^{+\sqrt{2W/m}} dU
.
\end{equation}
    The diamagnetic flow 
$\mbox{\boldmath$\Gamma$}_{\perp  1} 
= n_0 {\bf u}_{\perp 1}$ and the parallel flow 
$\Gamma_\parallel$ 
are of the first order in $\delta = \rho/L$ while 
 $\mbox{\boldmath$\Gamma$}_{\perp  2}^{\rm (rl)} 
= n_0 {\bf u}_{\perp 2}^{\rm (rl)}$ is of the second order. 
   The collisional particle conservation law, 
$\int d^3 v \, C^L (f) = 0$, is used to obtain  
Eq.~(\ref{continuity_eq}). 
  Also, it should be noted 
that the boundary condition,  
$({\bf V}_{\rm gc}^{\rm (rl)} )_\perp (\mu = 0 ) = 
({\bf V}_{\rm gc}^{\rm (rl)} )_\perp 
( U = \pm \sqrt{2W/m} ) = 0$,  
is used for deriving Eq.~(\ref{continuity_eq}) as well as 
the energy and parallel momentum balance equations 
[see Eqs.~(\ref{energy_balance}) and (\ref{para_mbalance})] 
from Eq.~(\ref{dke_rl}). 
   We find that the flux surface average of the left-hand side of 
Eq.~(\ref{continuity_eq}) automatically vanishes so that no particle 
source is required for obtaining the stationary solution. 
 Thus, the radially local approximation presented here has self-consistency 
with neglecting the radial transport that causes variation in the 
surface-averaged particles' number [see Eq.~(\ref{s-particle-balance})].

   Next, we multiply Eq.~(\ref{dke_rl}) with  
$(W - 5 T / 2)$ and take its velocity-space integral to 
derive 
\begin{equation}
\label{energy_balance}
\nabla \cdot 
( q_\parallel {\bf b} 
+ 
{\bf q}_{\perp  1} 
+ 
{\bf q}_{\perp  2}^{\rm (rl)} ) 
=  
Q
,
\end{equation}
where the parallel and perpendicular heat fluxes are 
given by 
\begin{eqnarray}
q_\parallel 
&  = &
\int d^3 v \, f 
\left( W - \frac{5}{2} T \right) U
,
\nonumber \\
{\bf q}_{\perp  1} 
&  = & 
\frac{5}{2} 
\frac{p_0 c X_2}{e B} 
\nabla s \times  {\bf b}
 ,
\nonumber \\
{\bf q}_{\perp  2}^{\rm (rl)} 
&  = &
\int d^3 v \, f 
\left( W - \frac{5}{2} T \right) 
%
%
({\bf V}_{\rm gc}^{\rm (rl)} )_\perp
%
%
,
\end{eqnarray}
and the collisional heat generation is defined by
\begin{equation}
\label{Q}
Q
= 
\int d^3 v \, C^L(f) W
. 
\end{equation}
    In Eq.~(\ref{energy_balance}),  
${\bf q}_{\perp  2}^{\rm (rl)}$ is the second-order flux
like $\mbox{\boldmath$\Gamma$}_{\perp  2}^{\rm (rl)}$ in 
Eq.~(\ref{continuity_eq}). 
   Taking the flux surface average of Eq.~(\ref{energy_balance}), we obtain 
\begin{equation}
\label{Q=0}
\langle Q \rangle = 0
, 
\end{equation}
which represents the collisional heat exchange balance that needs to be 
satisfied in the stationary state. 
%
%
Unequal temperatures $T_{a0} \neq T_{b0}$ can occur 
in the case of  $m_a/m_b \ll 1$ or $m_a/m_b \gg 1$  
where 
the characteristic time of 
the collisional thermal equilibration between the species $a$ and $b$ is 
much longer than 
the 90$^\circ$ scattering times due to like-species collisions 
characterized by $\tau_{aa}$ and $\tau_{bb}$. 
Then, $C_{ab}( f_{a0}, f_{b0} )$ does not vanish even for the local Maxwellian 
distribution functions $f_{a0}$ and $f_{b0}$ given by Eq.~(\ref{Maxwellian}) 
and it describes the above-mentioned slow collisional thermal equilibration 
although the linearized collision operator $C^L$ used for the  Eq.~(\ref{dke_rl2}) 
does not include this equilibrium part of the collision term.  
However, the heat generation $Q_{ab}$, which is defined by Eq.~(\ref{Q}) with the linearized operator $C_{ab}^L$ for collisions between different species $a$ and $b$, 
generally remains nonzero (even for the case of $T_{a0} = T_{b0}$). 
Therefore, Eq.~(\ref{Q=0}), 
which is rewritten as $\left\langle Q_a \right\rangle 
\equiv \sum_{b \neq a} \left\langle Q_{ab}\right\rangle = 0$ 
(recall $Q_{aa} \equiv 0$), is considered to be the physically reasonable 
condition that should be satisfied in the {\it multi-species} stationary state 
of the radially local model without requiring additional heat source or sink. 
%
%

Multiplying Eq.~(\ref{dke_rl}) with $m U$ and taking its velocity-space integral 
give the parallel momentum balance equation, 
\begin{equation}
\label{para_mbalance}
{\bf b} \cdot [ 
\nabla p_1 + \nabla \cdot 
( \mbox{\boldmath$\pi$}_1 
+ \mbox{\boldmath$\pi$}_2^{\rm (rl)} ) 
]
=  
n_0 e B  
\frac{\langle B E_\parallel \rangle}{ 
\langle B^2 \rangle}
+ F_\parallel
\end{equation}
where the first-order pressure $p_1$ and the viscosity tensors 
$\mbox{\boldmath$\pi$}_1$ 
and $\mbox{\boldmath$\pi$}_2^{\rm (rl)}$ are defined by
\begin{eqnarray}
\label{p1pi1pi2}
p_1
&  = &
\frac{2}{3}
\int d^3 v \, f  W 
,
\nonumber \\
\mbox{\boldmath$\pi$}_1 
&  = & 
\int d^3 v \, f  ( m  U^2 - \mu B )
\left(
{\bf b} {\bf b} - \frac{1}{3} {\bf I}
\right)
 ,
\nonumber \\
\mbox{\boldmath$\pi$}_2^{\rm (rl)} 
&  = &
\int d^3 v \, f \, m U
\left(
%
%
({\bf V}_{\rm gc}^{\rm (rl)} )_\perp
%
%
{\bf b}
+  {\bf b}
%
%
({\bf V}_{\rm gc}^{\rm (rl)} )_\perp
%
%
\right)
,
\end{eqnarray}
and the parallel friction force is given by
\begin{equation}
F_\parallel
= 
\int d^3 v \, C^L (f) m U
. 
\end{equation}
    The first-order viscosity tensor $\mbox{\boldmath$\pi$}_1$ 
is written in the form of the traceless part of  the CGL pressure tensor as 
$\mbox{\boldmath$\pi$}_1  
= ( p_\parallel - p_\perp) ( {\bf b} {\bf b} - \frac{1}{3} {\bf I})$,  
where $p_\parallel$ and $p_\perp$ represent 
the parallel and perpendicular pressures, respectively. 
   On the other hand, the second-order 
viscosity tensor $\mbox{\boldmath$\pi$}_2^{\rm (rl)}$, which 
is given by the correlation between the parallel velocity 
$U$ and the perpendicular drift velocity 
%
%
$
({\bf V}_{\rm gc}^{\rm (rl)} )_\perp
$
%
%
, 
can not be written in the CGL form.  
  We now multiply Eq.~(\ref{para_mbalance}) with
the magnetic-field strength $B$ and take its magnetic-surface average to 
derive 
\begin{equation}
\langle {\bf B} \cdot [ 
\nabla \cdot 
( \mbox{\boldmath$\pi$}_1 
+ \mbox{\boldmath$\pi$}_2^{\rm (rl)} ) 
]
\rangle 
=  
n_0 e \langle B E_\parallel \rangle
+ \langle B F_\parallel \rangle
,
\end{equation}
   which is used later to derive an alternative expression for 
the neoclassical particle flux.

The radial neoclassical particle flux is written as  
\begin{eqnarray}
\label{Gncl}
\Gamma^{\rm ncl}
& = & 
\left\langle
\int d^3 v \,
f V_{\rm gc}^s
\right\rangle
\nonumber \\
& = & 
 \frac{c}{e}
\left\langle
 \frac{\nabla s}{B} \cdot 
[{\bf b} \times 
(\nabla p_1 + \nabla \cdot 
\mbox{\boldmath$\pi$}_1 )]
\right\rangle
\nonumber \\
& = & 
 \frac{c}{e \chi' }
\left\langle
 \frac{\partial {\bf x}}{\partial \zeta} \cdot 
(\nabla p_1 + \nabla \cdot 
\mbox{\boldmath$\pi$}_1 )
\right\rangle
\nonumber \\
&  & \mbox{}
-  \frac{c B_\zeta}{e \chi' }
\left\langle
 \frac{\bf b}{B} \cdot 
(\nabla p_1 + \nabla \cdot 
\mbox{\boldmath$\pi$}_1 )
\right\rangle,
\end{eqnarray}
where $V_{\rm gc}^s = {\bf V}_{\rm gc} \cdot \nabla s$ is 
given by Eq.~(\ref{vgcs}). 
  Derivation of Eq.~(\ref{Gncl}) uses the following formula, 
\begin{equation}
\label{formula_B}
\chi' \frac{\nabla s \times {\bf b}}{B}
=
 \frac{\partial {\bf x}}{\partial \zeta}
 -
\frac{B_\zeta}{B} {\bf b}
, 
\end{equation}
and the Boozer coordinates $(s, \theta, \zeta)$,~\cite{Boozer} 
for which the 
contravariant poloidal and toroidal components, $B_\theta$ and $B_\zeta$, 
of the magnetic field ${\bf B}$ are flux-surface 
functions. 
   We see from Eq.~(\ref{Gncl}) that the neoclassical particle flux is caused by 
the spatial gradients of the first-order pressure and viscosity tensor. 
   It can be shown that the second-order 
viscosity tensor $\mbox{\boldmath$\pi$}_2^{\rm (rl)}$ defined in 
Eq.~(\ref{p1pi1pi2}) satisfies 
\begin{equation}
\label{formula_pi2a}
\left\langle \frac{\nabla s}{B} \cdot 
\left[{\bf b} \times (
\nabla \cdot 
\mbox{\boldmath$\pi$}_2^{\rm (rl)} ) 
\right]
\right\rangle 
=  
0
\end{equation}
and 
\begin{equation}
\label{formula_pi2b}
\left\langle \frac{\bf b}{B} \cdot (
\nabla \cdot 
\mbox{\boldmath$\pi$}_2^{\rm (rl)} ) 
\right\rangle 
=  
0
. 
\end{equation}
%
%
%
In deriving Eqs.~(\ref{formula_pi2a}) and (\ref{formula_pi2b}), 
it is convenient to write
$\mbox{\boldmath$\pi$}_2^{\rm (rl)}
= A ( {\bf B} {\bf w} + {\bf w} {\bf B} )$ 
with ${\bf w} \equiv ({\bf b} \times \nabla s) / B$. 
Then, we find 
$
{\bf w} \cdot (\nabla \cdot 
\mbox{\boldmath$\pi$}_2^{\rm (rl)} )  = 
w^2 {\bf B} \cdot \nabla A + A ( 
{\bf B} \cdot \nabla {\bf w} \cdot {\bf w}
- {\bf w} \cdot \nabla {\bf w} \cdot {\bf B})
=
w^2 {\bf B} \cdot \nabla A - A \nabla s \cdot 
( \nabla \times {\bf w})
=  \nabla \cdot ( A \nabla s \times {\bf w}  )
$ 
and 
$
( {\bf b}/B ) \cdot (\nabla \cdot 
\mbox{\boldmath$\pi$}_2^{\rm (rl)} )
= \nabla \cdot (A {\bf w}) + A ({\bf w} \cdot \nabla {\bf B} 
\cdot {\bf b} 
- {\bf b} \cdot \nabla {\bf B} \cdot {\bf w} ) / B 
= \nabla \cdot (A {\bf w}) + A \nabla s 
\cdot (\nabla \times {\bf B})/B^2
= \nabla \cdot (A {\bf w})
$, 
which lead to Eqs.~(\ref{formula_pi2a}) and (\ref{formula_pi2b}), 
respectively. 
%
%
     Then, using Eqs.~(\ref{formula_B})--(\ref{formula_pi2b}), we 
also have
\begin{equation}
\label{formula_pi2c}
 \left\langle \frac{\partial {\bf x}}{\partial \zeta} \cdot (
\nabla \cdot 
\mbox{\boldmath$\pi$}_2^{\rm (rl)} ) 
\right\rangle 
=  
0
. 
\end{equation}
It is found from Eq.~(\ref{Gncl}) 
and Eqs.~(\ref{formula_pi2a})--(\ref{formula_pi2c}) 
that the second-order 
viscosity tensor $\mbox{\boldmath$\pi$}_2^{\rm (rl)}$ 
in the radially local approximation cannot contribute to 
the neoclassical transport like 
the first-order pressure $p_1$ and viscosity tensor 
$\mbox{\boldmath$\pi$}_1$. 

We now use Eqs.~(\ref{para_mbalance}) and (\ref{formula_pi2b}) 
to rewrite the expression of the radial neoclassical 
particle flux in Eq.~(\ref{Gncl}) as
\begin{eqnarray}
\label{Gnclb}
\Gamma^{\rm ncl}
& = & 
 \frac{c}{e \chi' }
\left\langle
 \frac{\partial {\bf x}}{\partial \zeta} \cdot 
(\nabla p_1 + \nabla \cdot 
\mbox{\boldmath$\pi$}_1 )]
\right\rangle
\nonumber \\
&  & \mbox{}
-  \frac{c B_\zeta }{e \chi' }
\left(
n_0 e
 \frac{\langle B E_\parallel \rangle}{\langle B^2 \rangle} 
+ 
\left\langle
\frac{F_\parallel}{B} 
\right\rangle
\right)
.
\end{eqnarray}
    Here, the first surface-averaged part on 
the right-hand of Eq.~(\ref{Gnclb}) represents 
the nonaxisymmetric part of the neoclassical 
particle flux,~\cite{Shaing,Sugama1996a}
\begin{equation}
\label{Gna}
\Gamma^{\rm na}
=
 \frac{c}{e \chi' }
\left\langle
 \frac{\partial {\bf x}}{\partial \zeta} \cdot 
(\nabla p_1 + \nabla \cdot 
\mbox{\boldmath$\pi$}_1 )]
\right\rangle
.
\end{equation}
    Then,  we find from Eqs.~(\ref{Gnclb}) and (\ref{Gna})
that the radial electric current is written as
\begin{equation}
\label{rcurrent}
\sum_a e_a \Gamma_a^{\rm ncl}
= \sum_a e_a \Gamma_a^{\rm na} 
,
\end{equation}
where the quasineutrality $\sum_a n_{a0} e_a = 0$ and 
the collisional momentum conservation 
$\sum_a F_{\parallel a} = 0$ are used. 

For axisymmetric and quasi-axisymmetric systems, 
we have 
\begin{equation}
\label{p1z}
\left\langle \frac{\partial p_1}{\partial \zeta} 
\right\rangle 
=  
\left\langle \frac{\partial {\bf x}}{\partial \zeta} \cdot (
\nabla \cdot 
\mbox{\boldmath$\pi$}_1 ) 
\right\rangle 
=  
0
,
\end{equation}
from which
\begin{equation}
\label{Gna0}
\Gamma^{\rm na} 
= 0
, 
\end{equation}
and 
\begin{equation}
\Gamma^{\rm ncl}
=
-  \frac{c B_\zeta }{e \chi' }
\left(
n_0 e
 \frac{\langle B E_\parallel \rangle}{\langle B^2 \rangle} 
+ 
\left\langle
\frac{F_\parallel}{B} 
\right\rangle
\right)
\end{equation}
are derived. 
    Here, the quasi-axisymmetry means that 
the magnetic field strength $B = |{\bf B}|$ is independent of the 
toroidal angle $\zeta$. 
%
%
For derivation of Eq.~(\ref{p1z}), 
the $\zeta$-independence of $B$ and 
$\sqrt{g} = (4 \pi^2)^{-1} (dV/ds) \langle B^2 \rangle / B^2$ 
in the Boozer coordinates~\cite{Sugama2002} is used 
[see also Eq.~(25) in Ref.~17]. 
%
%
    It is confirmed from Eqs.~(\ref{rcurrent}) and (\ref{Gna0}) 
that the solution $f$ of the radially local drift kinetic equation 
shown in Eq.~(\ref{dke_rl}) or (\ref{dke_rl2}) 
gives the neoclassical particle fluxes which satisfy 
the ambipolarity condition, 
\begin{equation}
\label{ambipolarity}
\sum_a e_a \Gamma_a^{\rm ncl} = 0
, 
\end{equation}
automatically in axisymmetric and 
quasi-axisymmetric systems.  
    The intrinsic ambipolarity can be proved in the same way 
for all other quasi-symmetric systems such as 
quasi-poloidally-symmetric and quasi-helically-symmetric 
systems.  

%
%
It is seen from Eqs.~(\ref{rcurrent0})--(\ref{pa2sz}) that 
the radial current is closely related to the toroidal viscosity or the 
radial transport of the toroidal momentum. 
As remarked after Eq.~(\ref{pa2sz}), 
the ambipolarity condition is not guaranteed on the second order in $\delta$ 
for axisymmetric systems without up-down symmetry (as well as 
quasi-axisymmetric systems without stellarator symmetry) because of  
the component $(\pi_{a2})_\zeta^s$ of the second-order non-CGL viscosity 
tensor. 
We also note that the radial neoclassical particle flux defined by Eq.~(\ref{Gncl}) 
is the second-order flux driven by the first-order CGL tensor which 
becomes a dominant part for nonaxisymmetric systems although it  
does not contain the third-order flux due to the second-order tensor. 
Therefore, Eq.~(\ref{ambipolarity}) should be interpreted to imply that 
the intrinsic ambipolarity condition 
for the axisymmetric case can be 
correctly treated only up to the first order 
by the present radially local approximation. 
  On the other hand, 
the second-order neoclassical 
radial flux $\langle (\pi_{a2})_\zeta^s \rangle$ 
of the toroidal momentum 
in the axisymmetric but up-down asymmetric case can also be evaluated 
using the solution $f$ of the radially local drift kinetic equation 
even without resort to the radially global model. 
This can be done by substituting the solution $f$ into the formula for 
the toroidal momentum transport flux given by 
Eq.~(18) in Ref.~17. 
[It is confirmed from Eqs.~(11) and (13) in Ref.~29 
that, if using the definition of $\mbox{\boldmath$\pi$}_2$ 
given by Eq.~(\ref{nupppf}) in the present work 
to evaluate $\langle (\pi_{a2})_\zeta^s \rangle$, only a part of 
the result from Eq.~(18) in Ref.~17 is reproduced.]   
%
%

\section{ENTROPY PRODUCTION RATE AND ONSAGER SYMMETRY 
ASSOCIATED WITH NEOCLASSICAL TRANSPORT EQUATIONS}

The neoclassical radial particle flux $\Gamma^{\rm ncl}$, 
heat flux $q^{\rm ncl}$, and 
parallel electric current 
$J_E =  \langle B J_\parallel \rangle / \langle B^2 \rangle^{1/2}$ 
are defined in terms of 
the solution $f$ of Eq.~(\ref{dke_rl}) or (\ref{dke_rl2}) by  
\begin{eqnarray}
\label{nclflux}
\Gamma_a^{\rm ncl}
& = & 
\left\langle \int d^3 v \, f_a V_{{\rm gc}\, a}^s \right\rangle
,
\nonumber \\
q_a^{\rm ncl}
& = & 
\left\langle \int d^3 v \, f_a V_{{\rm gc}\, a}^s 
\left( W - \frac{5}{2} T_a \right) \right\rangle
,
\nonumber \\
J_E
& = & 
\langle B^2 \rangle^{-1/2}
\sum_a e_a 
\left\langle \int d^3 v \, f_a U \right\rangle 
,
\end{eqnarray}
where the subscript $a$ denotes the particle species. 
   The linearized collision operator in Eq.~(\ref{dke_rl}) 
for the species $a$ is defined in terms of 
the bilinear operator $C_{ab}$ 
for collisions between the species $a$ and $b$ 
by  
\begin{equation}
\sum_b 
[ C_{ab}( f_a, F_{b0}) + C_{ab} (F_{a0}, f_b) ]
. 
\end{equation}
Here,  $C_{ab}^T (f_a) \equiv C_{ab}(f_a, F_{b0})$ and 
$C_{ab}^F (f_b) \equiv C_{ab} (F_{a0}, f_b)$ are called 
test- and field-particle collision operators, respectively, and 
they satisfy the adjointness relations,~\cite{Sugama2009,RHH}
\begin{eqnarray}
\label{adjoint}
\int d^3 v \frac{f_a}{F_{a0}} C_{ab}^T (g_a)
& = & 
\int d^3 v \frac{g_a}{F_{a0}} C_{ab}^T (f_a)
,
\nonumber \\
T_a \int d^3 v \frac{f_a}{F_{a0}} C_{ab}^F (g_b) 
& = & 
T_b \int d^3 v \frac{g_b}{F_{b0}} 
%
%
C_{ba}^F (f_a) 
%
, 
\end{eqnarray}
and 
Boltzmann's H-theorem,~\cite{Sugama2009,RHH}
\begin{eqnarray}
\label{Htheorem}
& & 
T_a \int d^3 v \frac{f_a}{F_{a0}} 
[ C_{ab}^T (f_a)  +  C_{ab}^F (f_b) ]
\nonumber \\ & & 
\mbox{} + 
T_b \int d^3 v \frac{f_b}{F_{b0}} 
[ C_{ba}^T (f_b)  +  C_{ba}^F (f_a) ]
\leq 0
.
\end{eqnarray}
   Strictly speaking, the adjointness relations and the H-theorem 
are rigorously 
satisfied by the linearized Landau collision operator 
only for the case of $T_a = T_b$ 
although they are still approximately valid even for $T_a \neq T_b$
when $(m_a/m_b)^{1/2}$ or $(m_b/m_a)^{1/2} (1 - T_b/T_a)$ 
are small enough.~\cite{Sugama2009}

   Since the drift kinetic equations 
for different particle species are coupled with each other 
due to the field particle collision operators, 
$f_a$ depends not only on thermodynamic forces 
$(X_{a1}, X_{a2}, X_E)$  but 
also on those for  $b\neq a$, $(X_{b1}, X_{b2})$. 
   Accordingly,  we find that 
$\Gamma_a^{\rm ncl}$, $q_a^{\rm ncl}$, and 
$J_E$ in Eq.~(\ref{nclflux}) are related to the thermodynamic forces 
through the neoclassical transport equations which are written as 
\begin{eqnarray}
\label{Labij}
\Gamma_a^{\rm ncl}
& = & 
\sum_b
( L_{ab}^{11} X_{b1} + L_{ab}^{12} X_{b2})
+ L_{aE}^1 X_E
,
\nonumber \\
q_a^{\rm ncl}/T_a
& = & 
\sum_b
( L_{ab}^{21} X_{b1} + L_{ab}^{22} X_{b2})
+ L_{aE}^2 X_E
,
\nonumber  \\
J_E
& = & 
\sum_b
( L_{Eb}^{1} X_{b1} + L_{Eb}^{2} X_{b2})
+ L_{EE} X_E
.
\end{eqnarray}
     Here, the neoclassical transport coefficients 
$(L_{ab}^{11},  L_{ab}^{12}, \cdots )$ are 
regarded as functions of the variables
$[E_s ( \equiv - d\Phi/ds), \, \nabla s \cdot \nabla B, 
\, \nabla s \cdot ({\bf b} \cdot \nabla {\bf b})]$
which characterize the perpendicular guiding center velocity 
$({\bf V}_{\rm gc}^{\rm (rl)})_\perp$ defined in 
Eq.~(\ref{Vperp}). 

We here examine the Onsager symmetry of the neoclassical transport 
coefficients. 
In order to prove the Onsager symmetry, 
the adjointness relations written in Eq.~(\ref{adjoint}) and 
the phase-space-volume conservation along the collisionless guiding
center orbit are required as shown in 
Refs.~19 and 29. 
However, in the radially local model based on  Eq.~(\ref{dke_rl}), 
the latter condition 
$\nabla \cdot {\bf V}_{\rm gc}^{\rm (rl)}  + \partial  (
dU/dt ) / \partial U = 0$ is broken so that 
the Onsager symmetry is not satisfied. 

As noted before Eq.~(\ref{dke_rl}), 
the Jacobian for  the phase-space coordinates $({\bf X}, W, U, \xi)$ 
is given by $1/m$. 
Here, we consider a modified Jacobian, 
\begin{equation}
\label{DW}
D_W  = [ 1 + d_* ({\bf X}, W, U)  ] / m
,
\end{equation}
which differs from the one mentioned above by 
the correction term $d_*$ of ${\cal O}(\delta)$ 
[see Eq.~(\ref{Ddt}) below]. 
   This term $d_*$ is determined by assuming that 
the Jacobian $D_W$ satisfies the conservation law of the phase-space 
volume element written as 
\begin{equation}
\label{DWt}
\nabla \cdot ( D_W  {\bf V}_{\rm gc}^{\rm (rl)} )  + 
\frac{\partial }{\partial U} \left(
D_W \frac{dU}{dt} 
\right)
=0
,
\end{equation}
where ${\bf V}_{\rm gc}^{\rm (rl)}$ and $dU/dt$ are 
given by Eqs.~(\ref{gu_eqs_rl}) and (\ref{Vperp}). 
   We can rewrite Eq.~(\ref{DWt}) as 
\begin{eqnarray}
\label{Ddt}
& & 
\left( {\bf V}_{\rm gc}^{\rm (rl)}  \cdot \nabla \theta 
\frac{\partial}{\partial \theta} + 
{\bf V}_{\rm gc}^{\rm (rl)}  \cdot \nabla \zeta 
\frac{\partial}{\partial \zeta}
+ \frac{dU}{dt} \frac{\partial }{\partial U} \right) 
\ln D_W 
\nonumber \\
& = &
\left( {\bf V}_{\rm gc}^{\rm (rl)}  \cdot \nabla \theta 
\frac{\partial}{\partial \theta} + 
{\bf V}_{\rm gc}^{\rm (rl)}  \cdot \nabla \zeta 
\frac{\partial}{\partial \zeta} + 
\frac{dU}{dt} \frac{\partial }{\partial U} \right) 
\ln (1 + d_*) 
\nonumber \\
& = &
-
\nabla \cdot {\bf V}_{\rm gc}^{\rm (rl)} 
- \frac{\partial }{\partial U} \left(
\frac{dU}{dt} 
\right)
. 
\end{eqnarray}
Noting that 
the last line of Eq.~(\ref{Ddt}) is of ${\cal O}(\delta)$, 
we can take the correction term $d_*$ as a small quantity of ${\cal O}(\delta)$. 
   The left-hand side of Eq.~(\ref{Ddt}) represents the derivative of 
$\ln D_W $ along the radially local guiding center orbit labeled by 
the  constant parameters $(s, W)$. 
   Assuming that the guiding center orbit ergodically covers 
the $(\theta, \zeta, U)$ space, 
$D_W = (1 + d_*)/m$ is determined by Eq.~(\ref{Ddt}) except for 
a factor that is an arbitrary function of $(s, W)$. 
   In order to uniquely specify $D_W = (1 + d_*)/m$,  
we impose another constraint, 
\begin{equation}
\label{dconstr}
\left\langle \int_{-\sqrt{2W/m}}^{+\sqrt{2W/m}} dU \, d_*
\right\rangle 
=
0
,
\end{equation}
where $\langle \cdots \rangle$ represents the flux-surface average 
defined in Eq.~(\ref{saverage}).
   Owing to the condition in Eq.~(\ref{dconstr}), 
$d_*$ is given as a small correction and it does not 
affect the surface-averaged velocity integral of the 
equilibrium distribution function $F_0$ as 
shown by 
\begin{equation}
\left\langle \int d^3 v \, (1+d_*) F_0
\right\rangle 
=
\left\langle \int d^3 v \, F_0
\right\rangle 
=
n_0
,
\end{equation}
where $\int d^3 v$ is given by Eq.~(\ref{intv}) and 
$F_0$ is the local Maxwellian defined in Eq.~(\ref{Maxwellian}) 
with the equilibrium density $n_0$ and temperature $T_0$ 
given as flux-surface functions. 

We next define another distribution function $f_*$ by
\begin{equation}
\label{f*}
f_* \equiv \frac{f}{1 + d_*}
,
\end{equation}
and use Eq.~(\ref{DWt}) to rewrite
Eq.~(\ref{dke_rl}) in terms of $f_*$ as 
\begin{eqnarray}
\label{dke_rl3}
{\cal V} f_*
&  = &
\frac{F_0}{T_0}
\left\{
V_{\rm gc}^s
\left[  X_1  +  X_2 
\left( \frac{W}{T_0} - \frac{5}{2} \right)
\right]
+ \frac{e U B}{\langle B^2 \rangle^{1/2}} X_E
\right\}
\nonumber \\ & & 
\mbox{}
+ C^L (f_*)
, 
\end{eqnarray}
where the differential operator ${\cal V}$ is defined by 
\begin{equation}
\label{calV}
{\cal V} 
 \equiv 
(1+d_*)
\left(
 {\bf V}_{\rm gc}^{\rm (rl)} \cdot \nabla 
+  \frac{d U }{d t}
\frac{\partial }{\partial U}
\right)
. 
\end{equation}
The collision term $C^L(f)$ in Eq.~(\ref{dke_rl}) is 
replaced by $C^L(f_*)$ in Eq.~(\ref{dke_rl3}) where the 
deviation of $C^L(f_*)$ from $C^L(f)$ is of ${\cal O}(\delta^2)$ and it is 
neglected. 
   It is shown from Eq.~(\ref{DWt}) that
the differential operator ${\cal V}$ satisfies the antisymmetry
relation,   
\begin{equation}
\label{antisym}
\left\langle \int d^3v \,
\alpha {\cal V} \beta \right\rangle
= 
-
\left\langle \int d^3v \,
\beta {\cal V} \alpha \right\rangle
,
\end{equation}
where $\alpha$ and $\beta$ are 
arbitrary smooth functions on the phase space. 

   Replacing $f$ with $f_*$ in Eq.~(\ref{nclflux}), 
we can define modified transport fluxes, 
$\Gamma_{*a}^{\rm ncl}$, $q_{*a}^{\rm ncl}$, and $J_{*E}$, 
the values of which agree with those of 
$\Gamma_a^{\rm ncl}$, $q_a^{\rm ncl}$, and $J_E$, 
respectively, to the lowest order in $\delta$ because 
$f_* = f [ 1 + {\cal O} (\delta) ]$. 
   Then,  substituting the solution $f_*$ of Eq.~(\ref{dke_rl3}) into 
the definitions of the modified transport fluxes, 
we can derive the neoclassical transport equations relating  
$(\Gamma_{*a}^{\rm ncl}, q_{*a}^{\rm ncl},  J_{*E})$ 
to $(X_{b1}, X_{b2}, X_E)$. 
These transport equations take the same forms as those in 
Eq.~(\ref{Labij}), and we use 
$(L_{*ab}^{11}, L_{*ab}^{12}, \cdots)$ to 
represent the modified transport coefficients which correspond to  
$(L_{ab}^{11}, L_{ab}^{12}, \cdots)$ in Eq.~(\ref{Labij}), 
respectively. 
   It is shown in the same way as in Sec.~III that 
no additional sources and/or sinks are required to obtain 
stationary particle and energy balances from Eq.~(\ref{dke_rl3})

We now multiply Eq.~(\ref{dke_rl3}) for particle species $a$ with $T_a f_{*a}/F_{a0}$ 
and take its velocity-space integral, flux-surface average, and summation over species. 
Then,  we obtain 
\begin{eqnarray}
\label{entropybalance}
& & 
\sum_a 
\left(
T_a \Gamma_{* a}^{\rm ncl} X_{a1} + q_{* a}^{\rm ncl} X_{a2}
\right)
+ J_{*E} X_E
\nonumber \\
& = &  
-
\sum_{a,b} T_a
\left\langle 
\int d^3v \, \frac{f_{*a}}{F_{a0}}
\left[ C_{ab}^T  (f_{*a}) + C_{ab}^F  (f_{*b})  \right] 
\right\rangle
\geq 0
, 
\hspace*{7mm}
\end{eqnarray}
where the inequality is due to the H-theorem given in Eq.~(\ref{Htheorem}). 
Equation~(\ref{entropybalance}) means 
that the neoclassical transport process is subject to 
the second law of thermodynamics:  
the summation of products between the transport fluxes and forces equals 
the entropy production rate expressed 
in terms of the linearized collision operator, which 
is positive definite.

    Since the differential operator ${\cal V}$ and the linearized collision 
operator $C^L$ satisfy the antisymmetry relation in Eq.~(\ref{antisym}) and the adjointness relations in Eq.~(\ref{adjoint}), respectively, 
we can use the same procedures as in Ref.~29 
to prove that 
the modified transport coefficients 
$(L_{*ab}^{11}, L_{*ab}^{12}, \cdots)$ obey
the Onsager symmetry relations 
written as 
\begin{eqnarray}
\label{onsager}
L_{*ab}^{ij} ( \mbox{\boldmath$\beta$} )
& = &
L_{*ba}^{ji} ( -\mbox{\boldmath$\beta$} )
\hspace*{5mm}
(i, j = 1, 2)
,
\nonumber \\
L_{*aE}^{i} ( \mbox{\boldmath$\beta$} )
& = & 
- L_{*Ea}^{i} ( -\mbox{\boldmath$\beta$} )
\hspace*{5mm}
(i = 1, 2)
,
\nonumber \\
L_{*EE} ( \mbox{\boldmath$\beta$} )
& = & 
L_{*EE} ( -\mbox{\boldmath$\beta$} )
, 
\end{eqnarray}
where
$\mbox{\boldmath$\beta$} \equiv [E_s, \, \nabla s \cdot \nabla B, 
\, \nabla s \cdot ({\bf b} \cdot \nabla {\bf b})]$ 
represent the variables associated with 
the perpendicular guiding center velocity 
$({\bf V}_{\rm gc}^{\rm (rl)})_\perp$ 
as explained after Eq.~(\ref{Labij}). 
   Note that the change from $\mbox{\boldmath$\beta$}$ to 
$-\mbox{\boldmath$\beta$}$ corresponds to turning 
$({\bf V}_{\rm gc}^{\rm (rl)})_\perp$ in the opposite direction.

  The positive definiteness and the Onsager symmetry shown in 
Eqs.~(\ref{entropybalance}) and (\ref{onsager}) for 
the neoclassical transport defined by the solution $f_*$ 
of Eq.~(\ref{dke_rl3}) 
do not hold for  the neoclassical fluxes 
$(\Gamma_{a}^{\rm ncl}, q_{a}^{\rm ncl},  J_{E})$ and 
the transport coefficients 
$(L_{ab}^{11}, L_{ab}^{12}, \cdots)$ in Eq.~(\ref{Labij})
derived from the solution $f$ of Eq.~(\ref{dke_rl}). 
    On the other hand, 
we also find that $\Gamma_{* a}^{\rm ncl}$ defined by $f_*$ 
is not written  
in the same form as in Eq.~(\ref{Gnclb}) 
because 
the parallel momentum balance equation derived from 
Eq.~(\ref{dke_rl3}) cannot be used exactly in the same way as 
in deriving Eq.~(\ref{Gnclb}) from Eq.~(\ref{Gncl}). 
       Thus, 
the intrinsic ambipolarity condition 
for axisymmetric and quasi-symmetric systems  
is slightly broken by 
the modified neoclassical particle fluxes $\Gamma_{* a}^{\rm ncl}$ 
obtained using $(L_{*ab}^{11}, L_{*ab}^{12}, L_{*aE}^1)$ 
while it is rigorously satisfied by 
$\Gamma_a^{\rm ncl}$ using 
$(L_{ab}^{11}, L_{ab}^{12}, L_{aE}^1)$ 
as shown in Sec.~III.

By the way, it can be shown in the same way as 
in Ref.~29 
that, for  axisymmetric systems with up-down symmetry and 
helical systems with stellarator symmetry, 
the neoclassical transport coefficients 
$(L_{*ab}^{11}, L_{*ab}^{12}, \cdots)$ satisfy 
the restricted forms of the Onsager symmetry relations, 
\begin{eqnarray}
\label{onsager2}
L_{*ab}^{ij} ( \mbox{\boldmath$\beta$} )
& = &
L_{*ab}^{ij} ( - \mbox{\boldmath$\beta$} )
= 
L_{*ba}^{ji} ( \mbox{\boldmath$\beta$} )
\hspace*{5mm}
(i, j = 1, 2)
,
\nonumber \\
L_{*aE}^{i} ( \mbox{\boldmath$\beta$} )
& = & 
- L_{*aE}^{i} ( - \mbox{\boldmath$\beta$} )
=
 L_{*Ea}^{i} (  \mbox{\boldmath$\beta$} )
\hspace*{5mm}
(i = 1, 2)
,
\nonumber \\
L_{*EE} ( \mbox{\boldmath$\beta$} )
& = & 
L_{*EE} ( -\mbox{\boldmath$\beta$} )
.
\end{eqnarray}

\section{CONCLUSIONS}

In this paper, a novel radially local approximation 
of the drift kinetic equation is presented. 
   The approximated guiding center equations, which are shown in 
Eq.~(\ref{gu_eqs_rl}), have no radial drift velocity component but 
they maintain the ${\bf E}\times {\bf B}$ drift and 
the component of the magnetic drift tangential to the flux surface. 
   In addition, they conserve the particle kinetic energy at the 
expense of the conservation of the magnetic moment. 
   Under this approximation, 
   a new drift kinetic equation is given by Eq.~(\ref{dke_rl}) 
in the conservative form, which has favorable properties for 
numerical simulation that any additional terms for particle and 
energy sources are unnecessary for obtaining stationary solutions. 
   Also, it is shown to satisfy the intrinsic ambipolarity condition for 
neoclassical particle fluxes in axisymmetric and quasi-symmetric 
toroidal systems. 
  Another radially local drift kinetic equation 
is presented in Eq.~(\ref{dke_rl3}),  
the solution of which 
equals that of Eq.~(\ref{gu_eqs_rl}) to the leading order in the expansion 
with respect to the drift ordering parameter $\delta$ defined by the ratio of 
the gyroradius to the equilibrium scale length. 
  The  positive definiteness of the entropy production due to 
the neoclassical transport fluxes  
and the Onsager symmetry of the neoclassical transport coefficients 
are rigorously guaranteed by 
the solution of Eq.~(\ref{dke_rl3}) although it does not exactly assure  
the intrinsic ambipolarity condition for 
neoclassical particle fluxes in axisymmetric and quasi-symmetric 
systems.   
Thus, Eqs.~(\ref{dke_rl}) and (\ref{dke_rl3}) each have favorable 
properties which are weakly broken in the other equation.  
To the lowest order in $\delta$,  the neoclassical transport fluxes 
derived from both solutions of  Eqs.~(\ref{dke_rl}) and (\ref{dke_rl3}) 
have the same values as each other, and 
no additional sources and/or sinks are required for those solutions 
to satisfy stationary particle and energy balances consistently.  
  Therefore,  both drift kinetic equations are considered to be practically 
useful for numerically evaluating the neoclassical transport fluxes with 
including effects of  the  ${\bf E}\times {\bf B}$ and 
magnetic drift motions tangential to the flux surface 
in the framework of the radially local approximation. 
   Numerical applications of the present local model are in progress 
and their results will be reported elsewhere.

\begin{acknowledgments}

This work was supported in part by NIFS/NINS under 
the Project of
Formation of International Network for 
Scientific Collaborations and in part by  
the NIFS Collaborative Research Programs 
(NIFS14KNTT026).   

\end{acknowledgments}




\end{document}